# Transverse localization of light and its dependence on the phase-front curvature of the input beam in a disordered optical waveguide lattice

S Ghosh[1*], B P Pal[1], R K Varshney[1] and G P Agrawal[2]

[1] Physics Department, Indian Institute of Technology Delhi, Hauz Khas, New Delhi 110016, India
[2] Institute of Optics, University of Rochester, Rochester, NY 14627, USA

Email: somiit@rediffmail.com*

**Abstract**

We investigate the influence of the phase-front curvature of an input light beam on the transverse localization of light by choosing an evanescently coupled disordered one-dimensional semi-infinite waveguide lattice as an example. Our numerical study reveals that a finite phase front curvature of the input beam indeed plays an important role and it could degrade the quality of light localization in a disordered dielectric structure. More specifically, a faster transition from ballistic mode of beam propagation due to diffraction to a characteristic localized state is observed in case of a continuous wave (CW) beam, whose phase-front is plane as compared to one having a curved phase front.

## 1. Introduction

The concept of transverse localization of light [1] in disordered, one-dimensional (1D) and two-dimensional (2D) discrete optical systems has attracted a great deal of interest in view of its underlying interesting physics, and potential novel applications [2, 3]. It is known that, with deliberately introduced disorder, light confinement occurs only in a plane perpendicular to the direction of light transport, both in temporarily realized [4] and permanently formed lattices [5]. With the ongoing intense research on photonic bandgap structures and discrete photonic systems, light localization has emerged as an intense field of contemporary research in the context of disordered optical structures [4-11]. In a recent study, we have shown that this phenomenon of light localization is independent of the input beam shape by taking the example of an evanescently coupled 1D disordered waveguide lattice [7]. Interactions between initially chosen transverse phase of the propagating beam and multiple scattering that takes place with propagation along the disordered sample could significantly change the interference property, which is the key to light localization and hence we expect that an input beam with a curved phase-front could play an important role in the localization of light. However, no study in this direction has been reported to date.

In this paper, we numerically investigate the effect of the input beam having a finite phase-front curvature on its localization in a disordered coupled waveguide lattice.

## 2. Numerical modeling of light localization in a disordered waveguide lattice

We consider an evanescently coupled waveguide lattice consisting of a large number ($N$) of unit cells, and in which all the waveguides spaced equally apart are buried inside a medium of constant refractive index $n_0$ [12,13]. The overall structure is homogeneous in the longitudinal ($z$) direction along which the optical beam is assumed to propagate, as shown in figure 1(a). The perturbation in refractive index $\Delta n(x)$ (over the uniform background of $n_0$) due to disorder in this 1D waveguide lattice is assumed to be of the form

$$\Delta n(x) = \Delta n_p (H(x) + C\delta(x)) \qquad (1)$$

here $C$ is a dimensionless constant, whose value governs the level/strength of disorder; the periodic function $H(x)$ takes the value 1 inside the higher-index regions and is zero otherwise; $\Delta n(x)$ consists of a deterministic periodic part $\Delta n_p$ of spatial period $\Lambda$ and a spatially periodic random component $\delta$ (uniformly distributed over a specified range varying from 0 to 1). This particular choice of randomly perturbed refractive indices in the high index as well as low index layers enables us to model the diagonal and off-diagonal disorders to study the localization of light [5]. In our modeling, we have ignored any disorder in spatial periodicity [7, 14, 15]. Wave propagation through the lattice is governed by standard scalar Helmholtz

equation, which under paraxial approximation could be written as

$$i\frac{\partial A}{\partial z} + \frac{1}{2k}\left(\frac{\partial^2 A}{\partial x^2}\right) + \frac{k}{n_0}\Delta n(x) A = 0 \quad (2)$$

where $A(x,z)$ is amplitude of an input CW optical beam having its electric-field as $E(x,z,t) = \text{Re}[A(x,z)e^{i(kz-\omega t)}]$; $k = n_0 \omega/c$. To study the effect of having a curved phase front of an input beam on the phenomena of transverse localization of light, the initial field amplitude of such an input Gaussian beam centered at $x_0$ is assumed to be of the form

$$A(x) = A_0 \exp\left[-(1-iB)((x-x_0)/\omega_0)^2\right] \quad (3)$$

where the parameter $B$ represents the phase-front curvature, and $\omega_0$ is the characteristic spot size of the Gaussian beam. In the case of a CW beam having a plane-phase front as the input, for which $B = 0$, Eqn. (2) yields localized exponential solutions [1, 7].

We solve Eqn. (2) with the scalar beam propagation method which we have implemented in Matlab® and consider input beams characterized with different values of the $B$-parameter. Details of the modeling methodology have been discussed in [7]. Our chosen waveguide array consists of 150 evanescently coupled waveguides, each of 7 μm width and separated from each other by 7 μm (i.e. center-to-center spacing is 14 μm). Such a waveguide array is realizable through laser inscription in glass for direct observation of localized light [14]. We deliberately choose a relatively small refractive index contrast along with a relatively long unit-cell period compared to the wavelength ($\Lambda \gg \lambda$) to ensure that the bandgap effects remain negligible.

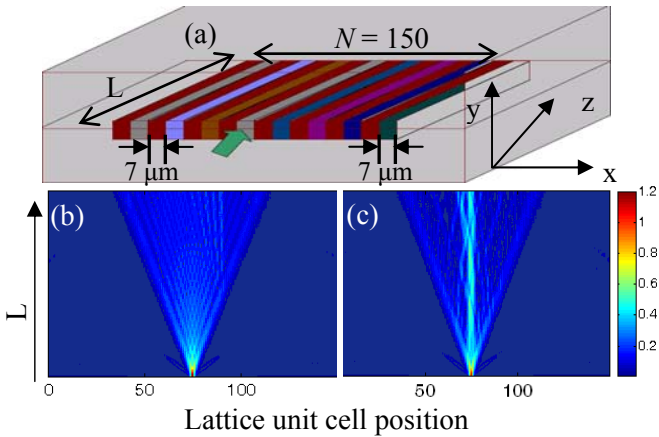

**Figure 1.** (Color online) a) Schematic of the refractive index disordered waveguide lattice with different colors indicating different refractive indices; b) ballistic mode of propagation through a perfectly ordered lattice; c) transition to a localized mode after propagation through a 20 mm ($L$) long 20% disordered lattice.

The value of $\Delta n_p$ was chosen to be 0.001 over and above the background material of refractive index $n_0 = 1.46$. To appreciate transverse localization of light in a disordered optical medium, we present in figure 1 results for propagation of an input CW beam with a plane phase front at an operating wavelength of 980 nm. Figure 1(a) depicts the array of waveguide lattices chosen for simulation study while figures 1(b) and 1(c) respectively correspond to $C = 0$ (indicating absence of disorder) and $C = 0.2$. The input beam is assumed to cover few lattice sites around the central unit cell at the input plane; chosen beam width $\omega_0$ (FWHM) was 12 μm (> width (7 μm) of an individual lattice site). From figure 1(c), a clear signature of transition to a localized state after an initial ballistic mode of propagation could be seen as the beam propagates along the length through the disordered waveguide lattice.

## 3. Results and Discussions

In order to investigate quantitatively the effect of having a finite phase front curvature of the input beam on its localization in a disordered medium, we have studied the beam dynamics for different lengths of the lattice and for different values of the $B$-parameter [incorporated in eq (3)]. A measure of the localization is assumed to be quantifiable through decrease in the average effective width ($\omega_{eff}$) (as defined in [7])

$$P \equiv \left[\int I(x,L)^2 dx\right] / \left[\int I(x,L) dx\right]^2$$
$$\omega_{eff} = \langle P \rangle^{-1} \quad (4)$$

of the propagating beam after including the statistical nature of the localization phenomenon in a finite system; where $\langle ... \rangle$ represents a statistical average over several realizations of the same level of disorder. After a certain propagation distance, depending upon the strength of disorder and lattice aspect ratio, $\omega_{eff}$ becomes almost unchanged with characteristic statistical fluctuations. It is worthwhile to point out that physically as we increase the value of $C$ parameter, the total number of localized eigen-states supported by a particular realization of the disordered lattice increases for a given length [5], which eventually favors a smaller $\omega_{eff}$ value. Naturally, a larger value of $\omega_{eff}$ in the localized regime as well as an increase in the localization length [4, 7] of an eigen-state in a disordered medium would imply degradation in the quality of localization. Results of such an effect due to a finite curvature of the input beam's phase front are illustrated in figure 2. This plot of $\omega_{eff}$ (ensemble averaged over 200 different realizations of a particular level of disorder and normalized with respect to $\omega_0$) with $L$ clearly reveals degradation in the quality of localization with propagation for a finite value of the parameter $B$ in the chosen disordered waveguide array.

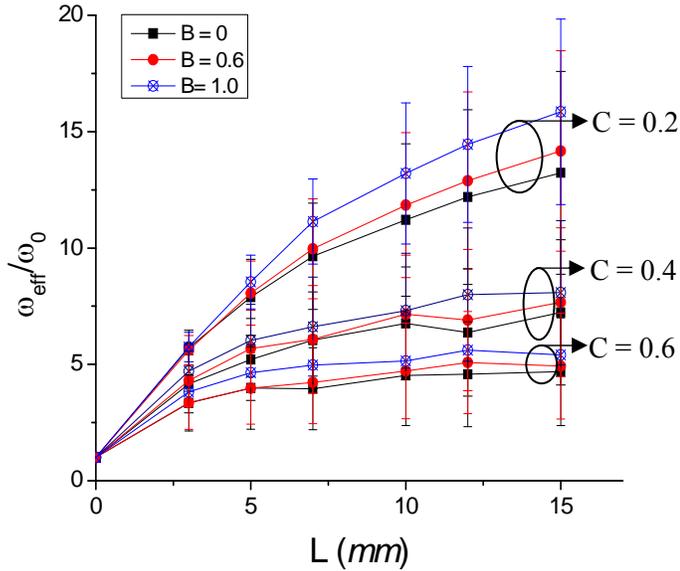

**Figure 2.** Variation in the ensemble averaged effective width ($\omega_{eff}$) of a Gaussian beam (of initial FWHM 12 μm) with propagation through the disordered waveguide lattice. The curves are labeled in terms of the disorder parameter $C$ for three different values of the phase front curvature parameter $B$. The error bars are the statistical standard deviations for the effective beam widths over 200 realizations.

For a waveguide lattice with a given strength of disorder, the localized eigen-channels supported by the lattice are given. A light beam having a finite input phase front curvature would preferably excite one of the extended (non-localized) natural eigen-states of a finite disordered lattice along with the localized states in few cases while performing the ensemble averaging over different realizations of the disorder. Hence, figure 2 essentially depicts the interplay between the strength of disorder and the phase front curvature of an input beam in the disordered lattice of finite length. As a sample result, we have plotted the beam dynamics for three different $B$-parameters when the values of $C$ are set at 0.2, 0.4 and 0.6, respectively. It may be noted that a similar trend was seen in our numerical simulations for negative values of the $B$-parameter.

To appreciate the above-mentioned degradation on the evolution of localization, in figure 3 we have plotted the ensemble-averaged (over 100 realizations) output intensity profiles from a 15 *mm* long disordered waveguide lattice of the above kind for $B = 0$ as well as a finite $B$-parameter (both positive and negative) when the level of disorder is set at $C = 0.2$. The plots in figure 3 correspond to an intermediate regime of propagation, in which both diffraction and localization are simultaneously present, before it attains a localized state. If we compare the plots corresponding to finite $B$ relative to the plot for $B = 0$, this particular trend also indicates that the finite B has a detrimental effect on simultaneous suppression of the ballistic side lobes to achieve characteristic localized tails. Hence, it can be concluded that an input beam with plane phase front would favor a faster transition to localization in comparison to a beam with a finite phase front curvature.

For a deeper appreciation of this effect on the degree of localization, we estimate the so called localization length ($l_C$) characteristic of a localized state in a particular disordered lattice [4]. To obtain $l_C$, we averaged 100 output intensity profiles for a given value of $C$ and then performed a three-point moving average to smoothen further the resulting profile as mentioned in [7]. As a particular localized state carries the signature of a corresponding disordered optical system, in general the state is not symmetric on either side of the peak. The corresponding variation of $l_C$ (fitted on either sides of the peak of the output intensity profile) for various $B$-parameters is shown in figure 4. These plots also confirm the advantage of a plane phase front over a curved one from localization point of view. It is obvious that larger the absolute value of the initial phase-front curvature, stronger would be the degradation in the quality of localization. These results should form useful guidelines while undertaking experimental studies on localization of light in a disordered dielectric.

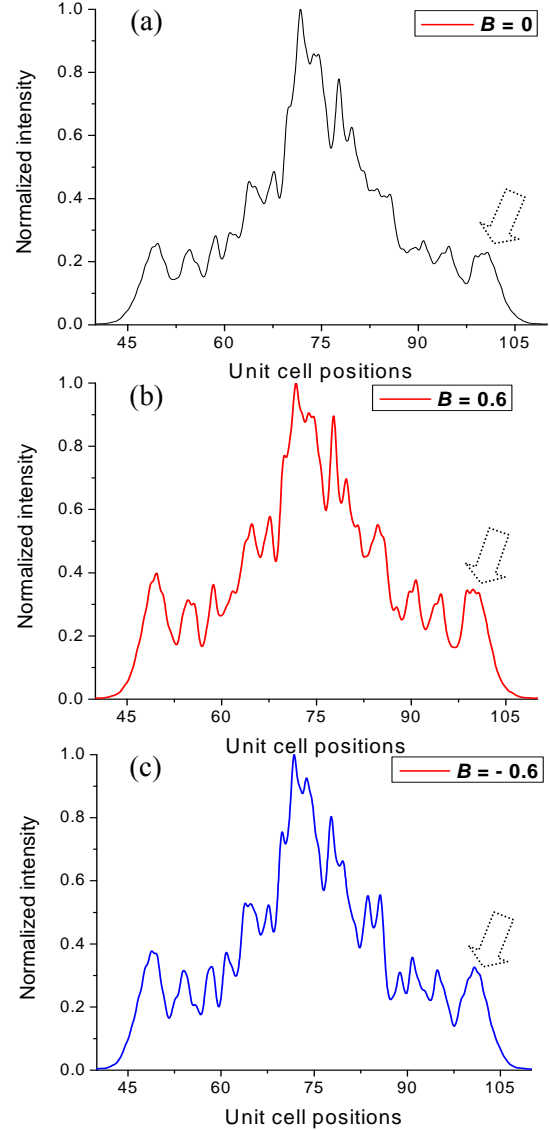

**Figure 3.** Comparison of building up of a central peak (a transition towards the localized state) and suppression of ballistic side lobes from figures (a), (b) and (c) respectively; clearly indicates a significant effect on localization transition due to initial curved phase front of the light beam. Side lobes are more prominent in the case of propagation of input beam with finite phase fronts.

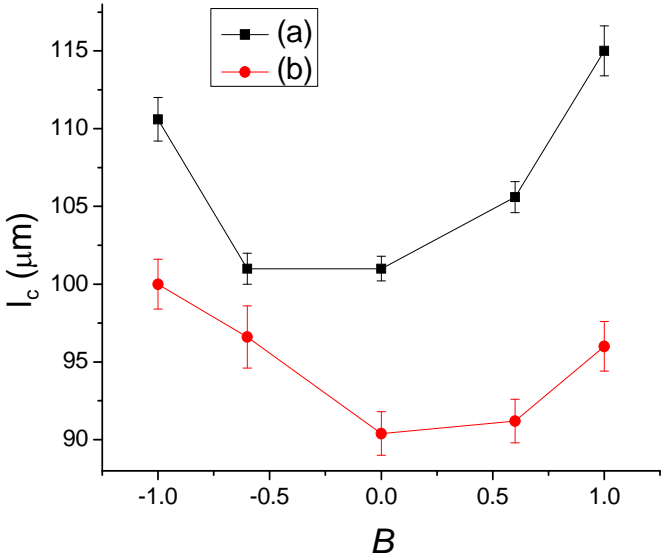

**Figure 4.** Localization lengths ($l_c$) on either side of a localized state have been plotted as a function of curvature parameter ($B$) for the same Gaussian input beam profile along 15 mm long lattice geometry of 60% disorder. Due to characteristic asymetry of a localized state around its peak, some difference in the variation of the localization length with $B$ on either side of peak of the profile is evident from (a) and (b). Bars show possible error encountered during curve fitting.

*3.1. Effect of input beam width*

Until now, we have chosen an input beam with a width $\omega_0$, which covers nearly four sites (high and low index) of the lattice to excite the eigen-modes supported by the same disordered waveguide lattice. Naturally, the input excites only those eigen-modes that are localized near the spatial location of excitation. In this subsection, we investigate the influence of the width of an input excitation while investigating the effect of a finite phase-front curvature on the phenomena of transverse localization of light. For this study we choose two different input beam widths that cover a single site and nearly 18 sites of the lattice. As before, the values of the $B$-parameter were chosen to be 0, ± 0.6, and the level of disorder was set at 20%. This choice of relatively small $C$ was to visualize the trend in variation of effective beam width with $B$ parameter for different strengths of disorder (as shown in Fig. 2), which clearly shows that the influence of a finite phase front curvature is more visible at a relatively weaker disorder. In figure 5, we have plotted the variation in $\omega_{eff}$ (normalized with respect to $\omega_0$) as a function of propagation length for the two different input beam widths. Figure 5a corresponds to the case when input beam width $\omega_0$ (3 μm) covers only one waveguide of the lattice; it could be seen that after propagation of only a few *mm* of the sample length, the $\omega_{eff}$ corresponding to the *B*-parameter (± 0.6) approaches to the nearly same final value, where as for *B* = 0 the beam evolved to a $\omega_{eff}$, which is much smaller relative to the previous value. Next, we choose a much broader input beam of $\omega_0$ (60 μm) that covers nearly 18 lattice sites and the corresponding propagation dynamics is shown in figure 5b. Interestingly to explain the behavior depicted in figure 5, we may need to analyze the results in terms of the transverse wave number ($k_\perp$) which is inversely proportional to the width ($\omega_0$) of the input beam [4]. This $k_\perp$ is large when the input beam is narrow and simultaneously if we introduce the phase front curvature as defined in Eqn. (3), the phase of the input beam strongly oscillates along the transverse profile of a localized mode. As a result, the input beam does not excite the localized modes efficiently. Apparently, the effect of this inefficient coupling as compare to the case of a plane phase front is likely to increase the $\omega_{eff}$ of the beam. In contrast, according to Eqn. (3), the phase is more likely to be homogeneous for an input beam with large $\omega_0$ and closer to the case of a plane phase front. Hence, we observe a relatively different propagation dynamics of the beam as compared to figure 5a. We have also verified that for an intermediate stage of the above two cases when we consider an input beam with $\omega_0$ (22 μm) which covers nearly 7 lattice sites follows the similar trend as shown in figures 5a & 5b. We may mention that the over-all trend of $\omega_{eff}$ variations for the three different cases (as shown in figure 5) remains almost unchanged with increasing the number of averaging and the nature of variations of individual curves establishes the fact that transverse wave number plays a key role while studying the effect of phase-front curvature. Whereas, the tall errors bars present in the plot carries the signature of the beam dynamics of the particular operating regime before entering the localized regime. Thus figure 5 presents the influence of the input beam width while studying the effect of a finite phase-front curvature and reveals that sufficiently broad input beam can downplay the adverse effect of input phase front curvature.

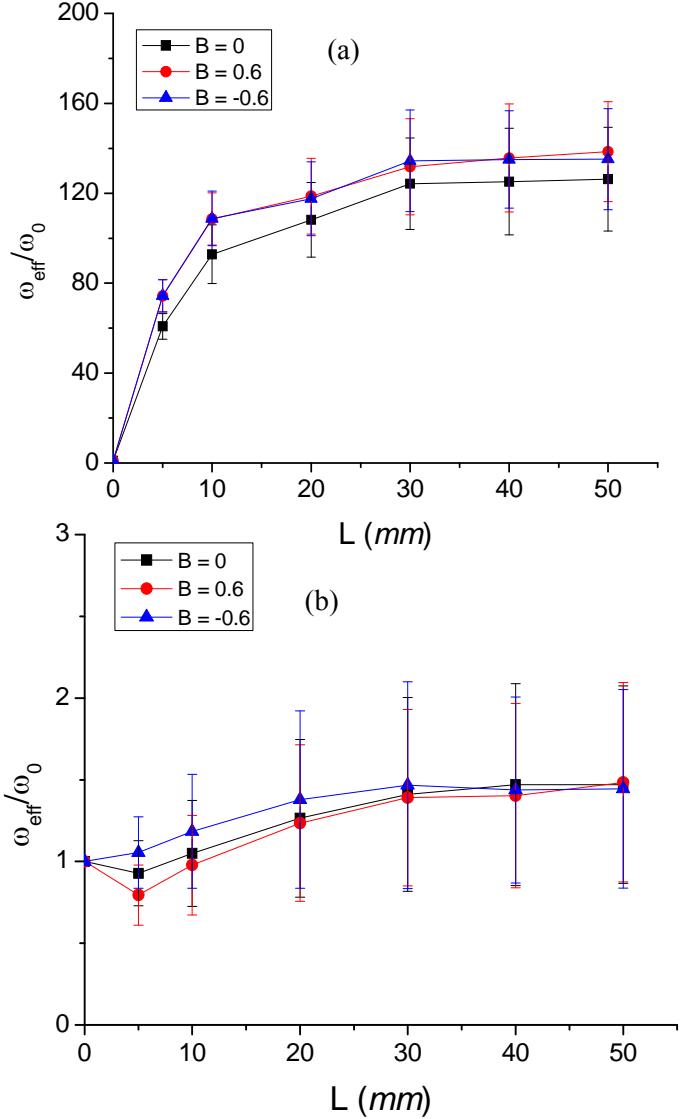

**Figure 5.** Variation of the ensemble averaged effective width ($\omega_{eff}$) of a Gaussian beam with two different input FWHMs with propagation through the waveguide lattice for $C = 0.2$. The error bars are the statistical standard deviations for the effective beam widths over 250 realizations. The chosen input widths cover; a) ~ one site of the lattice, b) ~ 18 sites of the lattice.

## 4. Conclusion

We have studied the significance of a finite initial phase front curvature of an optical beam in the context of transverse localization of light in a 1D coupled disordered waveguide lattice. The results of our extensive numerical simulations reveal that on an average, presence of an initial phase front curvature tends to degrade the effect of transverse localization. These results should be of interest in designing experiments related to study of transverse localization of light in a disordered medium. Also, an appropriate choice of the initial phase front can enable us to control the flow of light inside the disordered discrete photonics structures. Hence, we envision that initial phase front of a light beam introduces a new degree of freedom to manipulate the confinement of light in disordered optical systems.

## Acknowledgement

This work relates to Department of the Navy Grant N62909-10-1-7141 issued by Office of Naval Research Global. The United States Government has royalty-free license throughout the world in all copyrightable material contained herein. Thanks are due to Ishwar Aggarwal of Naval Research Laboratory Washington for interesting discussions.